\documentclass{article}
\usepackage{spconf,amsmath,graphicx}
\usepackage{amssymb}
\usepackage{hyperref}
\usepackage{algorithm,algorithmic}
\usepackage{pgfplots}

\title{Compressive Time Delay Estimation using Interpolation}
\name{Karsten Fyhn$^*$, Marco F. Duarte$^\dagger$ and S\o ren Holdt Jensen$^*$\thanks{E-mails: \{kfn,shj\}@es.aau.dk, mduarte@ecs.umass.edu. This work is supported by The Danish Council for Strategic Research under grant number 09-067056 and by a EliteForsk travel scholarship under grant number 11-116371.}}
\address{$^*$Dept. of Electronic Systems, Aalborg University, Denmark. \\
$^\dagger$Dept. of Electrical and Computer Engineering, University of Massachusetts Amherst, USA.
}

\usepackage{bm} 

\newcommand{\figref}[1]{Fig.~\ref{#1}}

\newcommand{\eqnref}[1]{(\ref{#1})}

\newcommand{\algoref}[1]{Algorithm~\ref{#1}}

\newcommand\va{\ensuremath{\mathbf{a}}}

\newcommand\vc{\ensuremath{\mathbf{c}}}

\newcommand\vf{\ensuremath{\mathbf{f}}}
\newcommand\vg{\ensuremath{\mathbf{g}}}
\newcommand\vh{\ensuremath{\mathbf{h}}}

\newcommand\vx{\ensuremath{\mathbf{x}}}
\newcommand\vy{\ensuremath{\mathbf{y}}}

\newcommand\vpsi{\ensuremath{\bm{\psi}}}
\newcommand\vphi{\ensuremath{\bm{\phi}}}

\newcommand\valpha{\ensuremath{\bm{\alpha}}}
\newcommand\vtau{\ensuremath{\bm{\tau}}}

\newcommand\mA{\ensuremath{\mathbf{A}}}
\newcommand\mB{\ensuremath{\mathbf{B}}}

\newcommand\mPsi{\ensuremath{\bm{\Psi}}}
\newcommand\mPhi{\ensuremath{\bm{\Phi}}}

\begin{document}
\maketitle

\begin{abstract}
Time delay estimation has long been an active area of research. In this work, we
show that compressive sensing with interpolation may be used to achieve good
estimation precision while lowering the sampling frequency.
We propose an Interpolating Band-Excluded Orthogonal Matching Pursuit algorithm that uses one
of two interpolation functions to estimate the time delay parameter. The
numerical results show that interpolation improves estimation precision and that
compressive sensing provides an elegant tradeoff that may lower the required
sampling frequency while still attaining a desired estimation performance.
\end{abstract}


\section{Introduction}
Time Delay Estimation (TDE) of one or more known signal waveforms from sampled
data is of interest in several fields such as radar, sonar, wireless communications,
audio, speech and medical signal processing. The classical problem is to obtain good
precision of the estimate while keeping the sampling frequency and computational
complexity low.

A popular method for TDE is to find the peak of the cross-correlation function.
As this must often be done on sampled data, the cross-correlation function is
sampled on a discrete grid. This corresponds to multiplying the received sampled
signal onto a dictionary of reference signals, each a delayed version of the
known waveform. The problem then reduces to finding the maximum nonzero
components in the cross-correlation vector. If a high estimation precision is desired,
this requires a high sampling rate, which may be costly to attain. Since the 
dimension of the sampled signal may often be much higher than the number of delays 
to be estimated in the signal, the cross-correlation vector may be assumed sparse 
and it is possible to lower the sampling rate by employing compressive sensing (CS)
\cite{Candes2006c,Donoho2006}. With CS, we seek to recover signals and
parameters from an under-determined system of linear equations by assuming
sparsity in a known dictionary. 
A common problem in CS is that the observed signals may not be sparsely
representable in the dictionary. This problem also occurs in TDE as the delay
parameter of the received waveforms is a continuous parameter. 

To remedy such lack of sparsity, we show that interpolation may be used in the CS framework to
improve estimation precision.
In this work, we focus on extending previous work on interpolation-based TDE and bridging it with
CS. Thereby, we attain good estimation precision while keeping the sampling
frequency low. We use a redundant dictionary of delayed waveforms as this
improves the estimation. Such redundancy, however, introduces coherence, and so
we use a coherence-inhibiting greedy algorithm for signal recovery. The
coherence-inhibition is similar to the model-based CS approach in
\cite{Baraniuk2010,Duarte2012,Duarte2013}. We compare the performance of four compressive delay
estimators: 1) an unmodified, coherence-inhibiting greedy algorithm that uses no
interpolation and therefore operates on a discrete grid, 2) an algorithm that
uses simple parabolic interpolation on the cross-correlation function, 3) an
algorithm that uses polar interpolation and 4) an algorithm that first reconstructs the full Nyquist- sampled signal using $\ell_1$-minimization as in classical CS \cite{Candes2006c,Donoho2006} and then estimates the delays using the MUSIC algorithm.
For all four estimators, 
we investigate their performance in terms of estimator precision with and without
Gaussian noise added. Furthermore, we briefly examine the computational complexity 
of the examined estimators. While we use a simple chirp signal in our numerical experiments, the
proposed approach may also be extended to more complex systems, e.g., symbol
synchronization in wireless communication systems.

\section{Problem Formulation and Background}
Let the received time-domain analog signal be defined as
\vskip-0.5cm
\begin{align}
	\label{eqn:ft}f(t;\valpha, \vtau)  &= \sum_{i=1}^K \alpha_i \cdot g(t-\tau_i) + n(t),
\end{align}
\vskip-0.2cm
\noindent
where $\valpha = \{\alpha_1, \alpha_2,\cdots,\alpha_K\}$ are the unknown signal
amplitudes, $\vtau = \{\tau_1, \tau_2,\cdots,\tau_K\}$ are the unknown signal
delays in time, $g(t)$ is a known signal waveform  and $n(t)$ is the noise.
The task of the estimation algorithm is then to estimate $\valpha$ and $\vtau$
from a sampled version of \eqnref{eqn:ft}. Depending on the bandwidth of $g(t)$,
the required sampling rate to estimate the delays to a sufficient precision may be high.
If we assume that only a few signal components are active, i.e. $K$ is small, we may use
CS to achieve the desired precision at a lower sampling rate. With a CS receiver, 
the received signal has the form $\vy = \Phi(f)$, 
where $\Phi(\cdot)$ represents a CS sampling structure such as Random
Demodulator \cite{Tropp2010} or Modulated Wideband Converter \cite{Mishali2010},
which takes as input a bandlimited signal such as $f(t)$ in \eqnref{eqn:ft}.
These CS sampling structures are designed so the sampling operation in the
analog domain is equivalent to a matrix-vector operation in the discrete domain, $\vy = \mPhi\vf$, 
where $\vf\in\mathbb{C}^{N}$ is the Nyquist sampled version of \eqnref{eqn:ft},
$\vy\in\mathbb{C}^{M}$ is the received signal and $\mPhi\in\mathbb{R}^{M\times
N}$ is the discrete equivalent of $\Phi(\cdot)$ with $N$ and $M$ the number of
Nyquist and measurement samples, respectively.

To enable reconstruction, CS requires a sparsifying dictionary
$\mPsi\in\mathbb{C}^{N\times N}$. In the case
of TDE, the dictionary is a circulant matrix, consisting of delayed waveforms:
\begin{align}
\nonumber\mPsi &= \begin{bmatrix}  \vpsi_0 & \vpsi_1 & \cdots & \vpsi_{N-1} \end{bmatrix} \\
        &= \begin{bmatrix}  g[0]    & g[N-1]    & \cdots & g[1] \\
                            g[1]    & g[0]      & \ddots & g[2] \\
                            \vdots  & \vdots    & \ddots & \vdots \\
                            g[N-1]  & g[N-2]    & \cdots & g[0] \end{bmatrix},
\end{align}
where $\vg=\begin{bmatrix} g[0] & g[1] & \cdots & g[N]\end{bmatrix}^T$ is the Nyquist-sampled version of $g(t)$ in
\eqnref{eqn:ft}. We use the term \emph{atom} to signify one column in this
dictionary, so that arbitrary signals are composed of atoms from the
dictionary. With this dictionary, the sampled cross-correlation function
may be obtained as $\hat{R}_f[n] = |\langle\vy,\vpsi_n\rangle|$. 
Since the delay parameter is continuous, the
received signal may not be perfectly representable by the sparsifying
dictionary, and the peak of the cross-correlation function then falls between its
sampled values. 

Prior work on this problem includes \cite{Jacques2008}, which uses a
gradient descent approach to approximate solutions off the grid for a generic
greedy algorithm. This approach is similar to one of the two algorithms proposed in 
\cite{Ekanadham2011b}, one using a first-order Taylor expansion, 
the other a form of polar interpolation. The authors show that polar interpolation 
outperforms Taylor expansion. In our work we extend upon the polar interpolation approach.
In \cite{Duarte2012,Duarte2013}, the authors use both a redundant dictionary with 
coherence rejection and second order polynomial interpolation to better estimate the 
solution. Similarly, in \cite{Fannjiang2012} the authors introduce algorithms that inhibit
coherent atoms in the recovery algorithms. Time delay estimation with CS has
been treated before in \cite{Gedalyahu2010}, where 
Estimation of Signal Parameters via Rotational Invariance Techniques (ESPRIT) is used to retrieve
the time delays after processing the signal with a specially designed filter bank. 
Their algorithm relies on periodic sequences of delayed signals and specially tailored 
analog filters that ensure stable inversion and is applicable to specific types of waveforms $g(t)$. 
In contrast, our approach processes the signal in the digital domain, 
provides a generic acquisition framework compatible with arbitrary waveforms $g(t)$, 
and does not require custom hardware filters or periodicity.
To compare our interpolation algorithms with a framework  
similar to that in \cite{Gedalyahu2010}, but for discrete, non-periodic sequences, 
we use an algorithm in the digital domain that first reconstructs the Nyquist-sampled 
signal using the Basis Pursuit (BP) algorithm \cite{Chen1998} for noise-less experiments 
or the Basis Pursuit Denoising (BPDN) algorithm \cite{Chen1998} for noisy experiments.
Then, we find the delay taps as $\hat{\vh} = \mPsi^\dagger \hat{\vf}$, where $\hat{\vf}$ 
is the reconstructed signal.
If all the delays are on the grid the delay tap vector, $\hat{\vh}$, has only $K$ active taps.
However, as is also seen for frequency sparse signals, $\hat{\vh}$ has side lobes when the delay
parameters are off the grid. A standard solution to solving this problem for 
frequency estimation is to use a high resolution algorithm, such as 
the Multiple Signal Classification (MUSIC) algorithm \cite{Stoica1997}. 
We therefore transform the time delay estimation problem to a frequency estimation problem 
by taking the inverse Fourier transform of $\hat{\vh}$ and use MUSIC to estimate the tap 
positions, which corresponds to the delay estimates. 
We term this algorithm TDE MUSIC.

\section{Interpolation in Time Delay Estimation}
Our contribution is bridging the work on CS and interpolation to improve
estimator precision in TDE while keeping the sampling frequency and computational complexity low.
This is achieved by proposing a new greedy algorithm with an interpolation step.
In each iteration of the algorithm, after finding the strongest correlating atom
(i.e. the largest absolute value of $\hat{R}_f[n]$), we
propose to use an interpolation function to improve the estimation precision:
\begin{align}
	\label{eqn:interpfunc}\tilde{\tau}_n = \mathrm{T}(\vy, \mPsi, i_n),
\end{align}
where $\tilde{\tau}_n$ is the new $n$th estimate of the delay, $\vy$ is the
received signal, $\mPsi$ is the dictionary and $i_n$ is the index for the atom
in the dictionary that features the strongest correlation with the signal.
There are many possible choices of interpolation functions. In this work we
compare two interpolation functions: second order polynomial and polar.

\emph{Polynomial interpolation} is a common method to increase the TDE
precision for sampled data \cite{Boucher1981,Jacovitti1993,Aiordachioaie2010}.
The simplest and most often used polynomial interpolation is fitting a parabola
around the correlation peak. In some cases, it is possible to improve the
estimation by using different polynomial interpolation techniques for different
problems, see, e.g., the references in \cite{Viola2005}.
In this work, we use the Direct Correlator estimator from \cite{Jacovitti1993}
for parabolic interpolation:
\begin{align*}
	\tau_i = -\frac{\Delta}{2}\frac{\hat{R}_f[(n+1)\Delta] - \hat{R}_f[(n-1)\Delta]}{\hat{R}_f[(n+1)\Delta] - 2\hat{R}_f[n\Delta] + \hat{R}_f[(n-1)\Delta]} + n\Delta,
\end{align*}
where $\tau_i$ is the delay to estimate, $\Delta$ is the spacing in time between
samples of the discrete cross-correlation function $\hat{R}_f$, and $n$ is the
index of the largest absolute entry in $\hat{R}_f$. 

\emph{Polar interpolation} is proposed in \cite{Ekanadham2011b} and obtains
improved estimates of off-the-grid atoms. This is done for shift-invariant
systems, where it can be assumed that the signal manifold of possible delayed
waveforms lies on a hypersphere \cite{Ekanadham2011b}, which is
also the case for time delay estimation. Such assumption is supported by the
fact that the magnitude $\|\vf\|_2$ of a delayed signal, $\vf$, is the same
regardless of the value of the parameter $\tau$; hence, the magnitude becomes
the radius of the modeling hypersphere.

Instead of interpolating using the cross-correlation function, polar
interpolation is based on the received signal itself and three atoms from the
dictionary; the strongest correlating atom, $\vpsi_{p}$, and its two adjacent
neighbours in the dictionary, $\vpsi_{p-1}$ and $\vpsi_{p+1}$. 
With these three vectors we may approximate a small part of the signal manifold's 
hypersphere with a circle arc. The interpolation
is obtained as follows; define the function $\vf_i$ as the sampled $i$th
signal component $\alpha_i \cdot g(t-\tau_i)$ in \eqnref{eqn:ft}. Then
$\vf_i$ may be approximated as
\begin{align}
	\label{eqn:polarintfc}\vf_i &\approx \vc^\mathrm{T} 
		\mA \begin{bmatrix} \vpsi_{p-1} \\ \vpsi_{p} \\ \vpsi_{p+1} \end{bmatrix},\qquad
    \vc = \begin{bmatrix} \alpha_i \\ \alpha_i r\cos\left(\frac{|\tau_i-n\Delta|\theta}{\Delta}\right) \\ \alpha_i r\sin\left(\frac{|\tau_i-n\Delta|\theta}{\Delta}\right) \end{bmatrix},\\	
    \label{eqn:polarintA} \mA &= \begin{bmatrix} 1 & r\cos(\theta) & -r\sin(\theta) \\ 1 & r & 0 \\ 1 & r\cos(\theta) & r\sin(\theta) \end{bmatrix}^{-1}, 
\end{align}
where $r=\|\vpsi_i\|_2$ is the magnitude of a signal waveform and radius of the
hypersphere, and $\theta$ is the angle between the vectors $\vpsi_{p}$ and
either $\vpsi_{p-1}$ or $\vpsi_{p+1}$. Notice that $r$ is identical for all
choices of $\tau$, hence the hypersphere assumption. 
In this formula, $\mA$ rotates the three $\vpsi$ vectors to form a new, general basis for the circle arc 
and $\vc$ scales the vectors in that basis to estimate the received signal.
Given a signal $\vf$ and
the atom in the dictionary that correlates the strongest with the signal
$\vpsi_p$, we may solve \eqnref{eqn:polarintfc} as a linear least squares problem with
$\vc$ as the unknown. From $\hat{\vc}=\{\hat{c}_1, \hat{c}_2,
\hat{c}_3\}$, we may obtain an estimate of $\tau_i$ by taking the inverse
tangent of $\hat{c}_3/\hat{c}_2$.


\section{Interpolating Band-Excluded Orthogonal Matching Pursuit}
We use a redundant, circulant dictionary which introduces coherence between atoms. 
To remedy the coherence effect we leverage the
Band-Excluded Orthogonal Matching Pursuit (BOMP) algorithm \cite{Fannjiang2012}
instead of classical orthogonal matching pursuit, as it inhibits atoms in the
recovered signal from being too closely spaced. In this work we compare: 1) the
original BOMP algorithm \cite{Fannjiang2012} and 2) Interpolating Band-Excluded
Orthogonal Matching Pursuit (IBOMP), which uses interpolation to estimate time
delays in between the sampling grid. The interpolation is done using parabolic or
polar interpolation. IBOMP is an extension to the BOMP algorithm
and is defined in \algoref{algo:ibomp}. First, the best correlating atom index
$i_n$ is found by generating a proxy for the sparse signal. This proxy is
trimmed using a band exclusion function \cite{Fannjiang2012}:
\begin{align}
	B_\eta(S) &= \cup_{k\in S} B_\eta(k),\\
	B_\eta(k) &= \{i \:|\: \mu(i,k) > \eta\},\;
	\mu(i,k) = |\langle\vpsi_i,\vpsi_k\rangle|,
\end{align}
where $\mu(i,k)$ is the coherence between two atoms in the dictionary,
$B_\eta(k)$ is the $\eta$-coherence band of the index $k$, and $B_\eta(S)$ is
the $\eta$-coherence band of the index set $S$. In this work, we set $\eta=0$,
as we assume that the signal waveforms are well spaced so that signal components
present in the received signal are orthogonal to each other.  Therefore, the
band-exclusion does not inhibit two pulses from interfering, but inhibits the
algorithm from finding the same pulse again due to a large remaining residual
$\vy_\textrm{res}$.  The selected atom is then input to the interpolation
function $\mathrm{T}(\vy, \mPsi, i_n)$, cf. \eqnref{eqn:interpfunc}, which finds
an estimate of the $n$th time delay, $\tilde{\tau}_n$. Using that time delay
estimate and the original parametric signal model, we create a new atom for a
signal dictionary, $\mB$, which is used to find the basis coefficients $\va$
using linear least squares. Finally, a new residual is calculated and $n$ and
$S$ are updated. When exiting the loop the signal is reconstructed. The stopping
criteria may be based on a noise floor estimate or, if the sparsity $K$ is
known, the loop is set to run $K$ times.
\begin{algorithm}[h!]
{\small
\caption{Interpolating Band-Excluded Orthogonal Matching Pursuit Algorithm (IBOMP)}
\label{algo:ibomp}
\begin{algorithmic}
	\STATE Input: Compressed signal $\vy$, interpolation function $\mathrm{T}(\vy, \mPsi, i_n)$, dictionary $\mPhi$ and measurement matrix $\mPsi$
    \vskip-0.4cm
	\STATE Output: Reconstructed signal $\tilde{\vx}$ and delay estimates $\tilde{\tau}_n$
	\STATE Initialize: $\vy_{\textrm{res}}=\vy$, $\mB=\emptyset$, $n=1$ and $S^n=\emptyset$
	\REPEAT 
		\STATE $i_n = \arg\max_i |\langle\vy_{\textrm{res}},\mPsi\vphi_i\rangle|,\:i \not\in B_0(S^{n-1})$
		\STATE $\tilde{\tau}_n = \mathrm{T}(\vy_{\textrm{res}}, \mPsi, i_n)$
		\STATE Include sampled version of $f(t-\tilde{\tau}_n)$ as new atom in $\mB$
		\STATE $\va = (\mPsi\mB)^\dagger \vy$
		\STATE $\vy_{\textrm{res}}=\vy-\mPsi\mB\va$
		\STATE $n = n+1,\;S^n = S^{n-1} \cup \{i_n\}$
	\UNTIL{stop-criterion = True}
	\STATE $\tilde{\vx} = \mB\va$
\end{algorithmic}
}
\end{algorithm} 

\section{Numerical Simulations}
To evaluate the proposed reconstruction algorithms, we have performed two
numerical experiments. The documentation and code for these experiments are made freely 
available at \url{http://www.sparsesampling.com/tde}, 
following the principle of Reproducible Research \cite{Vandewalle2009}.

For the numerical experiments, we let $g(t)$ in \eqnref{eqn:ft} be a chirp
signal defined as
\begin{align}
    g(t) &= \frac{1}{\sqrt{\mathcal{E}_g}} \cdot e^{j2\pi (f_0+\frac{\Delta f}{2T}(t-T/2))(t-T/2)} \cdot p(t), \\
    p(t) &= \left\{\begin{matrix} \frac{T}{2}(1 + \cos(2\pi (t-T/2)/T)), & t \in (0,T) \\ 0, & \text{otherwise} \end{matrix}\right.,
\end{align}
where $f_0=1$MHz is the center frequency, $\Delta f=40$MHz is the sweeped frequency, 
and $T=1\mu$s is the duration of the chirp in time. The chirp is limited in
time by a raised cosine pulse and normalized to unit energy. We assume
well-spaced pulses so that no two pulses overlap. Each signal is composed of
$K=3$ pulses, with $K$ known to the algorithms.

We perform Monte Carlo experiments and repeat each experiment $1000$ times to
get an average result. In each experiment, we generate a time signal by sampling
the pulse function in \eqnref{eqn:ft} $N=500$ times with a sampling frequency of $f_s=50$MHz. 
This sampling rate ensures that the corresponding bandwidth of the signal contains more than $99\%$ of its energy. The real and imaginary part of each $\alpha_i$ are drawn from a 
uniform distribution between $-10$ and $10$ and enforced to have a minimum absolute value of $1$.
The delays, $\tau$, are drawn from a uniform distribution between $0$ and 
$\frac{N-1}{f_s}-T=8.98\mu\text{s}$. 
For the CS measurement matrix we choose a Random Demodulator matrix \cite{Tropp2010}, 
$\mPsi\in\{-1,0,1\}^{M\times N}$. We set $M=\kappa N$, where $\kappa\in[0,1)$ is the CS 
subsampling rate. We evaluate the performance of the four estimators by computing the time delay
mean squared error ($\tau$-MSE) between the true and estimated value of the time delay. This corresponds to the sample variance of the estimators and is a measure of estimator precision. 

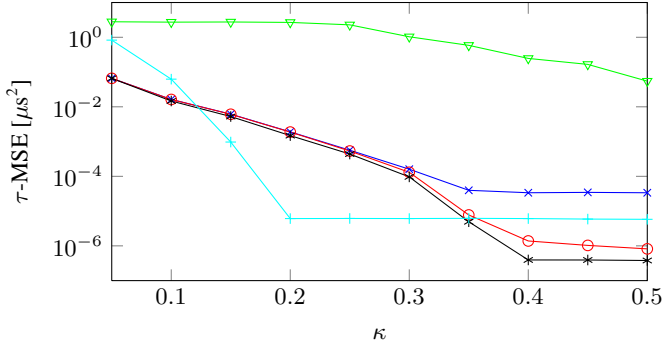
\begin{figure}[t]
	{\small
	\centering
	\newlength\figureheight 
	\newlength\figurewidth 
	\setlength\figureheight{3.7cm} 
	\setlength\figurewidth{0.4\textwidth}
%
%
%

\definecolor{mycolor1}{rgb}{0,1,1}

\begin{tikzpicture}

\begin{semilogyaxis}[%
width=\figurewidth,
height=\figureheight,
scale only axis,
xmin=0.05, xmax=0.5,
xlabel={$\kappa$},
ymin=1e-07, ymax=10,
yminorticks=true,
ylabel={$\tau\text{-MSE [}\mu{}\text{s}^\text{2}\text{]}$}]
\addplot [
color=blue,
solid,
mark=x,
mark options={solid},
forget plot
]
coordinates{
 (0.05,0.06592158019342)(0.1,0.016294395226275)(0.15,0.0061460431645886)(0.2,0.00190254742255535)(0.25,0.000564367603453209)(0.3,0.00016038045967065)(0.35,3.96200513635165e-05)(0.4,3.37177822368442e-05)(0.45,3.45342834156683e-05)(0.5,3.36924737086598e-05) 
};
\addplot [
color=red,
solid,
mark=o,
mark options={solid},
forget plot
]
coordinates{
 (0.05,0.0659640899582563)(0.1,0.0162609790030831)(0.15,0.00609853502598067)(0.2,0.00187730120347092)(0.25,0.000534871849110254)(0.3,0.000130634392382442)(0.35,7.73765574145862e-06)(0.4,1.38236705429322e-06)(0.45,1.03034093918379e-06)(0.5,8.18857507970851e-07) 
};
\addplot [
color=black,
solid,
mark=asterisk,
mark options={solid},
forget plot
]
coordinates{
 (0.05,0.0657566307484632)(0.1,0.0146990487865998)(0.15,0.00525161740167626)(0.2,0.00148485473034869)(0.25,0.000438117552018202)(0.3,9.64242337826378e-05)(0.35,4.93833452172222e-06)(0.4,3.93162859967526e-07)(0.45,3.91982658643585e-07)(0.5,3.79163329428366e-07) 
};
\addplot [
color=mycolor1,
solid,
mark=+,
mark options={solid},
forget plot
]
coordinates{
 (0.05,0.828276176836346)(0.1,0.062451989499192)(0.15,0.000969172722018644)(0.2,6.09477233827743e-06)(0.25,6.1372580448388e-06)(0.3,6.10461799304449e-06)(0.35,6.18276072511957e-06)(0.4,6.09447111312711e-06)(0.45,5.89168488754429e-06)(0.5,5.79917414643797e-06) 
};
\addplot [
color=green,
solid,
mark=triangle,
mark options={solid,,rotate=180},
forget plot
]
coordinates{
 (0.05,2.79524677164442)(0.1,2.72493073129445)(0.15,2.75501785703367)(0.2,2.67648460045949)(0.25,2.28012032440399)(0.3,1.033816951972)(0.35,0.586159264081825)(0.4,0.246466708688625)(0.45,0.166029875648566)(0.5,0.0547380496098545) 
};
\end{semilogyaxis}
\end{tikzpicture}%
   	\vskip-0.4cm
    \caption{$\tau$-MSE or variance versus subsampling ratio $\kappa$. See \figref{fig:Fig2} for legend.} 
	\label{fig:Fig1}
	}
\end{figure}

For the first experiment, we assume a noise-free signal. We perform this
experiment with a range of subsampling ratios $\kappa$. Figure \ref{fig:Fig1} shows our results.
All four estimators allow for subsampling while maintaining good estimation
precision. TDE MUSIC performs best for low $\kappa$, while the interpolation algorithms perform best as $\kappa$ increases. The Polar IBOMP algorithm is the best performing interpolation algorithm.
To compare the proposed CS algorithms with a method that does not use CS, 
we also show an algorithm that directly downsamples the signal by a factor of $N/M$ 
and then estimates the delays using the MUSIC algorithm.
As expected it does not attain the same estimation precision
as the CS-based algorithms, due to aliasing.
As a note on the y-axis and as validation for our implementation, notice that
BOMP converges to approximately $0.3\cdot 10^{-4}(\mu\text{s})^2$ which is to be expected since
the time delay between columns of the dictionary $\mPhi$ is
$1/50\text{MHz}=0.02\mu\text{s}$. Therefore the average error squared is
$\left(0.02/4\right)^2=2.5\cdot 10^{-5}(\mu\text{s})^2$, which corresponds well with
the numerical results.

For the second experiment we include additive white Gaussian measurement noise
in the signal model. We fix $\kappa=0.5$ and vary the signal-to-noise ratio
(SNR) from $-5$ to $30$~dB. The noise is generated by calculating the noise
power as the desired SNR divided by the measurement power, which is found as
the squared $\ell_2$-norm of the measurement vector.
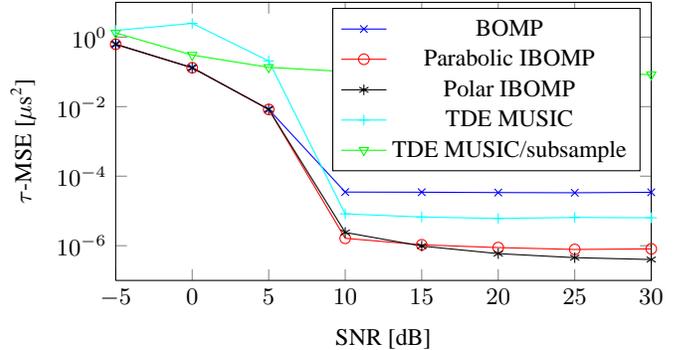
\begin{figure}[t]
	{\small
	\centering
	\setlength\figureheight{3.7cm} 
	\setlength\figurewidth{0.4\textwidth}
%
%
%

\definecolor{mycolor1}{rgb}{0,1,1}

\begin{tikzpicture}

\begin{semilogyaxis}[%
width=\figurewidth,
height=\figureheight,
scale only axis,
xmin=-5, xmax=30,
xlabel={SNR [dB]},
ymin=1e-07, ymax=10,
yminorticks=true,
ylabel={$\tau\text{-MSE [}\mu{}\text{s}^\text{2}\text{]}$},
legend style={draw=black,fill=white,align=left}]
\addplot [
color=blue,
solid,
mark=x,
mark options={solid}
]
coordinates{
 (-5,0.622147155545573)(0,0.131194952649987)(5,0.00834008480851809)(10,3.48813562012162e-05)(15,3.45364184244095e-05)(20,3.3756863884204e-05)(25,3.33100116804017e-05)(30,3.432327783795e-05) 
};
\addlegendentry{BOMP};

\addplot [
color=red,
solid,
mark=o,
mark options={solid}
]
coordinates{
 (-5,0.621949002301974)(0,0.131251104795091)(5,0.00830774498124563)(10,1.62974032178717e-06)(15,1.06015099616078e-06)(20,8.82073494543595e-07)(25,7.78802936289973e-07)(30,8.08838489280377e-07) 
};
\addlegendentry{Parabolic IBOMP};

\addplot [
color=black,
solid,
mark=asterisk,
mark options={solid}
]
coordinates{
 (-5,0.621974705627991)(0,0.131485480486364)(5,0.00837405166494928)(10,2.40316246018335e-06)(15,9.74201322793071e-07)(20,5.90637162820917e-07)(25,4.52320268329487e-07)(30,3.99789530799635e-07) 
};
\addlegendentry{Polar IBOMP};

\addplot [
color=mycolor1,
solid,
mark=+,
mark options={solid}
]
coordinates{
 (-5,1.5523894917478)(0,2.48592087417967)(5,0.206426443870316)(10,8.22753594552624e-06)(15,6.68200241555059e-06)(20,6.07207493195457e-06)(25,6.48696297607039e-06)(30,6.34805826017866e-06) 
};
\addlegendentry{TDE MUSIC};

\addplot [
color=green,
solid,
mark=triangle,
mark options={solid,,rotate=180}
]
coordinates{
 (-5,1.31989308100477)(0,0.304759360708025)(5,0.134876957503946)(10,0.102296953778845)(15,0.105234441581798)(20,0.109920671057395)(25,0.0967625871561128)(30,0.0831026282286283) 
};
\addlegendentry{TDE MUSIC/subsample};

\end{semilogyaxis}
\end{tikzpicture}%
   	\vskip-0.4cm
	\caption{$\tau$-MSE or variance versus SNR in decibel for $\kappa=0.5$.}
	\label{fig:Fig2}
	}
\end{figure} 
\figref{fig:Fig2} shows that the five algorithms are affected by noise,
but as the SNR increases they converge to the same performance as in the noiseless
case for $\kappa=0.5$ in \figref{fig:Fig1}.

\section{Computational Complexity}
\vskip-0.2cm
To compare the computational complexity of the four estimators, we first look at TDE MUSIC.
This algorithm consists of two parts: an $\ell_1$ minimization problem, in which the solution to the Newton system has complexity $O(N^3)$ \cite{Nesterov2004}, where $N$ is the signal length, and the MUSIC algorithm, which is dominated by computing the Singular Value Decomposition (SVD), has complexity of $O(N^3)$ for a square matrix \cite{Trefethen1997}.
For the algorithms based on BOMP the most significant term is to find the 
amplitude coefficients, $\va$ in \algoref{algo:ibomp}, 
where the pseudo-inverse is found by solving a linear least squares (LS) problem, 
which has cost \cite{Trefethen1997}:
\vskip-0.4cm
\begin{align}
    \text{Cost}_{\text{LS with SVD}} \sim 2NK^2 + 11K^3 = 2NK^2 + 11K^3. 
    \label{eqn:lssvd}
\end{align}
\vskip-0.1cm
\noindent This is for the last iteration of the BOMP algorithm, as this is the most costly 
when all $K$ atoms are used in the LS problem. 
As we assume $K\ll N$, the term $O(NK^2)$ dominates.

The interpolation in BOMP also adds some complexity.
For the polar interpolation, it is possible to compute the radius and angle
beforehand and create a dictionary of all possible sets of $\vpsi_{p-1},
\vpsi_{p}, \vpsi_{p+1}$ for all $p$. This dictionary may be multiplied with the
$\mA$ matrix in \eqnref{eqn:polarintA} beforehand and the remaining steps are
then dominated by finding the LS solution to \eqnref{eqn:polarintfc}, $\hat{\vc}$. 
If the complex LS problem is solved using a
complex SVD, the total cost is $O(N)$ and derived as
follows: For real-valued matrices/vectors the cost is identical to \eqnref{eqn:lssvd} 
where $K=3$ is the number of variables, i.e. $\text{Cost}_{\text{LS with SVD}} \sim 18N + 297$. 
As complex arithmetic can be reduced to real arithmetic, we simply say the
computational complexity is $O(N)$.

For the parabolic interpolation, we estimate the delay based on the cross
correlation function evaluated in three places. The BOMP algorithm has already
found these three values and the estimation complexity is therefore independent
of the problem size, i.e. $O(1)$. Thus, it is clear that the parabolic
interpolation is less computationally demanding than the polar interpolation and that the greedy algorithms are much less computationally demanding than TDE MUSIC.

\section{Conclusion}
We have compared four time delay estimators and shown
that all four methods are compatible with CS. Out of the four, TDE MUSIC performs the best for low values of $\kappa$, while IBOMP with
polar interpolation obtains the best performance with higher values of $\kappa$. We also show
that the interpolating greedy algorithms are less computationally demanding than TDE MUSIC.
\bibliographystyle{IEEEbib}
\bibliography{my_bibtex}

\begin{thebibliography}{10}

\bibitem{Candes2006c}
E.~J. Cand\`{e}s et~al.,
\newblock ``Stable signal recovery from incomplete and inaccurate
  measurements,''
\newblock {\em Comm. Pure Appl. Math.}, vol. 59, no. 8, pp. 1207--1223, 2006.

\bibitem{Donoho2006}
D.~L. Donoho,
\newblock ``Compressed sensing,''
\newblock {\em IEEE Trans. Inf. Theory}, vol. 52, no. 4, pp. 1289--1306, Apr.
  2006.

\bibitem{Baraniuk2010}
R.G. Baraniuk et~al.,
\newblock ``Model-based compressive sensing,''
\newblock {\em IEEE Trans. Inf. Theory}, vol. 56, no. 4, pp. 1982--2001, Apr.
  2010.

\bibitem{Duarte2012}
M.~F. Duarte,
\newblock ``Localization and bearing estimation via structured sparsity
  models,''
\newblock in {\em IEEE Statistical Signal Processing Workshop (SSP)}, 2012.

\bibitem{Duarte2013}
M.~F. Duarte and R.~G. Baraniuk,
\newblock ``Spectral compressive sensing,''
\newblock {\em Appl. Comput. Harmon. Anal.}, Accepted for publication.

\bibitem{Tropp2010}
J.~A. Tropp et~al.,
\newblock ``{Beyond Nyquist: Efficient sampling of sparse bandlimited
  signals},''
\newblock {\em IEEE Trans. Inf. Theory}, vol. 56, no. 1, pp. 520--544, Jan.
  2010.

\bibitem{Mishali2010}
M.~Mishali and Y.C. Eldar,
\newblock ``{From theory to practice: Sub-Nyquist sampling of sparse wideband
  analog signals},''
\newblock {\em IEEE J. Sel. Topics Signal Process.}, vol. 4, no. 2, pp.
  375--391, Apr. 2010.

\bibitem{Jacques2008}
L.~Jacques and C.~De~Vleeschouwer,
\newblock ``A geometrical study of matching pursuit parametrization,''
\newblock {\em IEEE Trans. Signal Process.}, vol. 56, no. 7, pp. 2835--2848,
  July 2008.

\bibitem{Ekanadham2011b}
C.~Ekanadham et~al.,
\newblock ``Recovery of sparse translation-invariant signals with continuous
  basis pursuit,''
\newblock {\em IEEE Trans. Signal Process.}, vol. 59, no. 10, pp. 4735--4744,
  Oct. 2011.

\bibitem{Fannjiang2012}
A.~Fannjiang and W.~Liao,
\newblock ``Coherence pattern-guided compressive sensing with unresolved
  grids,''
\newblock {\em SIAM J. Img. Sci.}, vol. 5, no. 1, pp. 179--202, Feb. 2012.

\bibitem{Gedalyahu2010}
K.~Gedalyahu and Y.C. Eldar,
\newblock ``Time delay estimation: Compressed sensing over an infinite union of
  subspaces,''
\newblock in {\em IEEE International Conference on Acoustics Speech and Signal
  Processing (ICASSP)}, Mar. 2010, pp. 3902--3905.

\bibitem{Chen1998}
S.~S. Chen et~al.,
\newblock ``Atomic decomposition by basis pursuit,''
\newblock {\em SIAM J. Sci. Comput.}, vol. 20, pp. 33--61, 1998.

\bibitem{Stoica1997}
P.~Stoica and R.~L. Moses,
\newblock {\em Introduction to spectral analysis},
\newblock Prentice Hall, Upper Saddle River, NJ, 1997.

\bibitem{Boucher1981}
R.~Boucher and J.~Hassab,
\newblock ``Analysis of discrete implementation of generalized cross
  correlator,''
\newblock {\em IEEE Trans. Acoust., Speech, Signal Process.}, vol. 29, no. 3,
  pp. 609--611, June 1981.

\bibitem{Jacovitti1993}
G.~Jacovitti and G.~Scarano,
\newblock ``Discrete time techniques for time delay estimation,''
\newblock {\em IEEE Trans. Signal Process.}, vol. 41, no. 2, pp. 525--533, Feb.
  1993.

\bibitem{Aiordachioaie2010}
D.~Aiordachioaie and V.~Nicolau,
\newblock ``On time delay estimation by evaluation of three time domain
  functions,''
\newblock in {\em 3rd International Symposium on Electrical and Electronics
  Engineering (ISEEE)}, Sept. 2010, pp. 281--286.

\bibitem{Viola2005}
F.~Viola and W.F. Walker,
\newblock ``A spline-based algorithm for continuous time-delay estimation using
  sampled data,''
\newblock {\em IEEE Trans. Ultrason., Ferroelectr., Freq. Control}, vol. 52,
  no. 1, pp. 80--93, Jan. 2005.

\bibitem{Vandewalle2009}
P.~Vandewalle et~al.,
\newblock ``{Reproducible research in signal processing [What, why, and
  how]},''
\newblock {\em IEEE Signal Process. Mag.}, vol. 26, no. 3, pp. 37--47, May
  2009.

\bibitem{Nesterov2004}
Y.~Nesterov,
\newblock {\em Introductory Lectures on Convex Optimization: A Basic Course},
\newblock Kluwer Academic Publishers, 2004.

\bibitem{Trefethen1997}
L.~N. Trefethen and D.~Bau,
\newblock {\em Numerical Linear Algebra},
\newblock Society for Industrial and Applied Mathematics, 1997.

\end{thebibliography}

\end{document}